\documentclass[aps,prb,twocolumn,floatfix,showpacs]{revtex4}

\usepackage{amssymb}

\usepackage{graphicx}
\usepackage{dcolumn}
\usepackage{bm}
\usepackage{color}

\begin{document}

\title{Inducing and Optimizing Magnetism in Graphene Nanomesh}
\date{\today}

\author{H.~X.~Yang}
\affiliation{SPINTEC, CEA/CNRS/UJF-Grenoble 1/Grenoble-INP, 38054 Grenoble, France}

\author{M.~Chshiev}
\email{mair.chshiev@cea.fr}
\affiliation{SPINTEC, CEA/CNRS/UJF-Grenoble 1/Grenoble-INP, 38054 Grenoble, France}

\author{D. W. Boukhvalov}
\affiliation{School of Computational Sciences, Korea Institute for Advanced Study (KIAS),
Hoegiro 87, Dongdaemun-Gu, Seoul, 130-722, Korean Republic}

\author{X.~Waintal}
\affiliation{SPSMS-INAC-CEA, 17 rue des Martyrs, 38054 Grenoble, France}

\author{S.~Roche}
\affiliation{Institut Catal\`a de Nanotecnologia (ICN) and CIN2, Campus UAB, E08193 Bellaterra, Barcelona, Spain} 
\affiliation{Institucio Catalana de Recerca i Estudis Avan\c cats (ICREA), 08010 Barcelona, Spain}

\begin{abstract}
Using first-principles calculations, we explore the electronic and magnetic properties of graphene nanomesh (GNM), a regular network of large vacancies, produced either by lithography or nanoimprint. When removing an equal number of A and B sites of the graphene bipartite lattice, the nanomesh made mostly of zigzag (armchair) type edges exhibit antiferromagnetic (spin unpolarized) states. In contrast, in situation of sublattice symmetry breaking, stable ferri(o)magnetic states are obtained. For hydrogen-passivated nanomesh, the formation energy is dramatically decreased, and ground state is found to strongly depend on the vacancies shape and size. For triangular shaped holes, the obtained net magnetic moments increase with the number difference of removed A and B sites in agreement with Lieb's theorem for even A+B. For odd A+B triangular meshes and all cases of non-triangular nanomeshes including the one with even A+B, Lieb's theorem does not hold anymore which can be partially attributed to introduction of armchair edges. In addition, large triangular shaped GNM could be as robust as non-triangular GNMs, providing possible solution to overcome one of crucial challenges for the sp-magnetism. Finally, significant exchange splitting values as large as $\sim 0.5$~eV can be obtained for highly asymmetric structures evidencing the potential of GNM for room temperature carbon based spintronics. These results demonstrate that a turn from 0-dimensional graphene nanoflakes throughout 1-dimensional graphene nanoribbons with zigzag edges to GNM breaks localization of unpaired electrons and provides deviation from the rules based on Lieb's theorem. Such delocalization of the electrons leads the switch of the ground state of system from antiferromagnetic narrow gap insulator discussed for graphene nanoribons to ferromagnetic or nonmagnetic metal.
\end{abstract}
\pacs{81.05.ue, 75.70.-i, 75.75.-c}

\maketitle

\section{INTRODUCTION}

Two-dimensional graphene has emerged as a natural candidate for developing "beyond CMOS" nanoelectronics~\cite{NovoselovScience, NovoselovNature, QHEGrNature, ConfinementGrScience, GeimNM,MaterToday}. In addition to the reported huge charge mobilities, the weak intrinsic spin-orbit  coupling in carbon-based $sp^{2}$ structures~\cite{SOC_C, SOC_Gr} could potentially allow for very large (micron long) spin diffusion lengths. These features, together with the other "semi-conductor like" properties of graphene, make graphene-based spintronic devices highly promising~\cite{Fert,Wolf} and have triggered a quest for controlling spin injection in graphene~\cite{SpinTransportGr,Han1,Han2,TYYang}. Many routes have been attempted to induce magnetism by proximity effect or inject spins from magnetic electrodes~\cite{FertAPL}.  Another, more intrinsic, possibility is  shaping the geometry of graphene by designing graphene nanoribbons with zigzag edges. This has been found to induce localized edge magnetic states which can serve as a conceptually new building block for spintronics~\cite{GNRzigzag, AFGr,  MagneticCorrelations,island}. 

The existence of intrinsic magnetism driven by atomic-scale defects (such as vacancies, chemisorbed species, grain boundaries,...) has been also suggested theoretically~\cite{MagIrradiationinduced,MagneticCorrelations,antidotsPRL, antidotsPRB,dotribbon,nanohole,tailor,graphitefilm}, but remains fiercely debated on the experimental side~\cite{EXPMAG}. It is indeed particularly difficult to achieve a precise experimental characterization of those defects, whereas the control of their density, positioning, or chemical reactivity seems an insurmountable challenge, jeopardizing a further use of magnetic properties in real devices. Additionally, the absence of a true energy gap in two-dimensional graphene limits the elaboration of active graphene-based devices and circuits with standard semiconductor technologies. 

Other route to make graphene magnetic is either chemisorption of odd number of adatoms or functional groups~\cite{MagneticCorrelations, clust}, or using magnetism on zigzag edges~\cite{GNRzigzag, YK, edgmag}. In the first case the stability of magnetic configurations at room temperature can be easily destroyed by the migration of adatoms with turning the system into nonmagnetic configuration~\cite{clust}. In contrast to the adatom based magnetism edge~\cite{edgmag} and vacancy~\cite{Krash} magnetism in graphite is stable at room temperature. But herewith localization of the magnetic moments on the edges provides formation of the AFM exchange interactions between two edges~\cite{GNRzigzag}. 

The case of graphene nanoribbons obey Lieb's theorem because the localized electrons on one edge belong to sublattice A and localized electrons from other edge to sublattice B. Magnetism on the edges of graphene nanoflakes is also described by this theorem~\cite{nanoflake}. Electron localization plays an important role in the different many-body effects on graphene edges~\cite{SKedges} and bulk graphene~\cite{SKbulk}. Understanding the nature of the electron localization and delocalization in graphene and related systems is necessary not only for control and manipulation of magnetism in studied compounds but also for the development of knowledge about systems with strongly correlated electrons. 

Graphene nanomesh (further GNM) is the intermediate compound between graphene nanoribbons with localized electrons on zigzag edges and perfect bulk graphene with delocalized electrons. The fabrication of GNM, using block copolymer lithography and offering versatility in varying periodicities and neck widths down to 5 nm~\cite{GNM}, could circumvent the hurdles. Indeed, such technique allows a scalable engineering of superlattices of large graphene vacancies whose density, shape and distribution can be controlled down to the nanoscale. Additionally, GNM-based field-effect transistors were shown to withstand current densities two orders of magnitudes larger than individual graphene nanoribbon devices, with comparable on/off ratio and easily tunable by varying the neck width. Moreover, in accordance to the well-established Lieb's theorem~\cite{spinGroundStateTheorem}, the possibility to control inner edge structures~\cite{advmat} of nanomesh vacancies could enable a true control of intrinsic magnetic properties.

In this paper, we present first-principles calculations of electronic and magnetic properties of graphene nanomesh, and found that by varying the shape, different types of intrinsic ferrimagnetic~\cite{comment} states can be obtained with clear identification of optimal conditions. Systematic studies of non-passivated and hydrogen-passivated GNM are achieved by varying the difference (\mbox{$\Delta_{AB}=|B-A|$}) between missing A and B sites of the underlying bipartite lattice and analyzing different hole geometries. For non-passivated GNM with $\Delta_{AB}=0$, stable non-magnetic states are found for armchair edge termination, while zigzag edge terminations result in antiferromagnetic ground states. These localized magnetic moments however vanish when all edge C-atoms are hydrogen-passivated.  In sharp contrast, when $\Delta_{AB}\not=0$, stable ferri(o)magnetic states are induced with net moment up to 4~$\mu_B$ (per 6 $\times$ 6 unit cell) originating from dangling bonds of edge atoms. 

Furthermore, for hydrogen-passivated GNM, the formation energy is dramatically decreased, and ground state is found to strongly depend on the vacancies shape and size. Our calculations reveal the existence of three magnetic regimes which depend on $\Delta_{AB}$: (i) highly magnetic GNMs obeying Lieb's theorem corresponding to triangular shaped holes with even A+B; (ii) GNMs with quenched magnetic state due to complete chemical bond reconstruction with $\Delta_{AB}=1$ and trivially nonmagnetic state with $\Delta_{AB}=0$; and (iii) GNMs following intermediate regime between magnetic and quenched magnetic states, i.e. triangular GNMs with odd A+B and more complicated structures including both even (e.g. sector shaped GNM) and odd (e.g. pentagon shaped GNM) A+B. We show that large triangular GNMs could be as robust as non-triangular GNMs providing possible solution to overcome one of crucial challenges for the sp-magnetism. Moreover, significant exchange splitting values as large as $\sim 0.5$~eV can be obtained for highly asymmetric structures evidencing the potential of graphene nanomesh for room temperature carbon based spintronics.

\section{METHODS}

First-principles calculations were performed using Vienna $ab ~initio$ simulation package (VASP)~\cite{vasp} based on density functional theory (DFT) with generalized gradient approximation (GGA) for exchange correlation potential. We have used projected augmented wave method (PAW) ~\cite{PAW} with the Perdew-Becke-Erzenhof (PBE) parametrization~\cite{PBE} potentials to describe the core electrons of carbon. Periodic 6 by 6 unit cells were used to simulate non-passivated GNM structures as shown in Fig.~\ref{fig1}, whereas periodic 8 by 8 unit cells were used to simulate H-passivated GNM structures. The kinetic cutoff energies for the plane wave basis set used to expand the Kohn-Sham orbitals were 520 eV for the self-consistent energy calculations. Methfessel-Paxton method~\cite{smear} is used with a broading width of 0.2 eV for the partial occupicancy smearing calculations. A 9$\times$9$\times$1 k-point mesh was sufficient to ensure good convergence in the total energy differences. The structural relaxations were performed ensuring that the Hellmann-Feynman forces acting on ions were less than 10$^{-3}$ eV/\AA.

\section{MODEL OF GRAPHENE NANOMESH}

\begin{figure}
\includegraphics[width=8 cm]{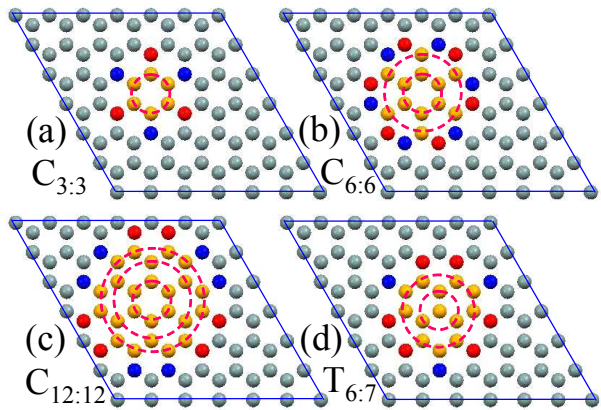}
\caption{(a)-(c): Schematics of the calculated crystalline structures for balanced non-passivated circular shaped C$_{3:3}$, C$_{6:6}$ and C$_{12:12}$ GNM structures, respectively; (d) the same for unbalanced non-passivated triangular shaped T$_{6:7}$ GNM structure. Edge carbon atoms are in blue and red color to represent A and B sites, respectively. For convenience, positions of removed atoms are indicated in orange.}
\label{fig1}

\end{figure}

A GNM can be formed by either removing atoms centered of a six ring structure [Fig.~\ref{fig1}(a,b,c)] or centered of one carbon atom [Fig.~\ref{fig1}(d)], which either lead to form GNM with balanced or unbalanced number of removed A and B sites. For the sake of clarity, we label those structures according to their shapes and put the number of removed A and B site atoms as a subscript. For instance, the structure of Fig.~\ref{fig1}(a) named C$_{3:3}$,  corresponds to a Circle hole shape GNM with 3:3 denoting 3~A and 3~B atoms removed from perfect graphene. The superscript H is used for hydrogen-passivated GNM. 

\section{RESULTS AND DISCUSSIONS}

For C$_{3:3}$ structure [Figure~\ref{fig1}(a)], we find that the configuration with opposite spin orientation between adjacent edge C atoms is energetically favored in comparison with the configuration with parallel spins between edge atoms of two sublattices represented by the blue and red color in Figure~\ref{fig1}. The total energy calculations reveal quite large magnetic interaction energies. For instance, the energy difference between ferromagnetic (FM) spin-polarized and paramagnetic (PM) state is found to be 0.129 eV per edge atom. The spin configuration is further stabilized by 0.093 eV per edge atom as a result of the antiferromagnetic (AF) coupling between neighboring atoms with magnetic moment of 0.48~$\mu_B$ per edge atom for each spin on each sublattice with opposite orientation. The magnetic moment is slightly larger than that of graphene nanoribbons which is $\sim$0.43~$\mu_B$~\cite{GNRzigzag}.

\begin{figure}
\leavevmode
\includegraphics[width=8 cm]{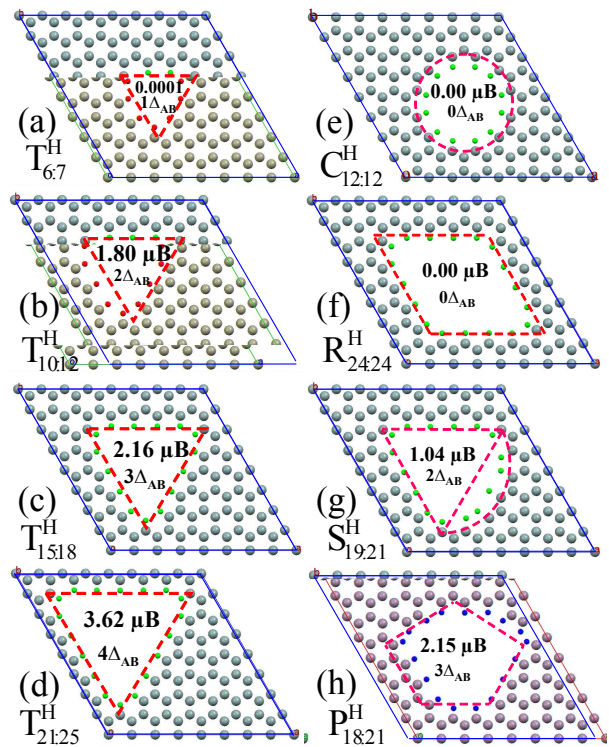}
\caption{H-passivated GNMs with triangular shapes: (a) T$^H_{6:7}$, (b)T$^H_{10:12}$, (c)T$^H_{15:18}$, (d) T$^H_{21:25}$; with circled shape (e) C$^H_{12:12}$; with rhombic shape (f) R$^H_{24:24}$; with sector shape (g) S$^H_{19:21}$; and with pentagon shape (h) P$^H_{18:21}$. The corresponding net magnetic moments for each structure are also indicated.}
\label{fig2}

\end{figure}

We now discuss the case of a 3-ring defect C$_{12:12}$ GNM [Fig.~\ref{fig1}(c)]. The total energy calculations show that the ground state is AF with a magnetic moment of 0.45~$\mu_B$ per edge atom for each spin on each sublattice with opposite orientation. The FM state is lower by 0.127 eV per edge atom compared to PM state, and the spin configuration is further stabilized by 0.107 eV per edge atom as a result of the AF coupling between neighboring atoms on different sublattice with opposite spin orientations~[see TABLE \ref{table:table I}].

\begin{table}
\caption {The number of removed atoms on A and B sites and their difference $\Delta_{AB}$. magnetic moment M($\mu_B$), total energies E (in eV) for ferrimagnetic(FMi), antiferromagnetic(AF) and nonmagnetic(NM) states and defect formation energies E$_{f}$ (in eV) for different GNM structures. The non-passivated C$_{3:3}$, C$_{6:6}$,C$_{12:12}$, and T$_{6:7}$ and passivated C$^H_{12:12}$ GNMs are calculated with 6$\times$6 unit cell, others are calculated with 8$\times$8 unit cell.}\label{table:table I}
	\centering
		\begin{tabular} {c  c c c c c c}
			\hline\hline
		     Structure          	 & $\Delta_{AB}$   & M($\mu_B$) &  & E$_{total}$ (eV) & & E$_{f}$(eV/C)\\
									   &            &                 & PM & AF & Ferri(o) & \\           
		\hline
			C$_{3:3}$ 	       	        & 0    & 0    & -590.414        &	\textbf  {-591.742} & (-591.185) & 2.81\\
	 		C$_{6:6}$ 	    	         & 0    & 0    & \textbf {-528.112}	  &	unstable & unstable & 2.10\\
	 		C$_{12:12}$ 	            & 0    & 0    & -406.137	  &	\textbf  {-408.953} & (-407.659) & 1.402 \\
	 		T$_{6:7}$ 	                 & 1    & 3.99 & -516.732 	    &	unstable & \textbf {-518.255} & 1.98 \\
											 &		&		&			&			&	(-517.630) & \\
										 	
			T$^H_{6:7}$ 	            & 1    & 10$^{-4}$ & -1089.046  & unstable & -1089.046 & 0.142\\
			T$^H_{10:12}$ 	            & 2    & 1.80 & -1015.387  & unstable  & \textbf {-1015.403} & 0.120\\
	 		T$^H_{15:18}$ 	           & 3    & 2.16 & -923.346  & unstable & \textbf {-923.365} & 0.103\\
	 		T$^H_{21:25}$ 	            & 4    & 3.62 & -813.104  & unstable & \textbf {-813.223} & 0.084\\	 	
	 		S$^H_{19:21}$ 	            & 2    & 1.04 & -862.346  & unstable  & \textbf {-862.348} & 0.082\\
			P$^H_{18:21}$	           & 3    & 2.15 & -872.125  & unstable & \textbf {-872.147} & 0.156\\
			C$^H_{3:3}$ 	            & 0    & 0    & \textbf {-627.565}	  &	unstable & unstable & 0.22 \\
	 		C$^H_{12:12}$ 	            & 0    & 0    & \textbf {-481.358}	  &	unstable & unstable & 0.078 \\
	 		R$^H_{24:24}$ 	            & 0    & 0    & \textbf {-795.267}  &   unstable & unstable & 0.070\\	
			P$'^H_{18:19}$	            & 1    & 10$^{-4}$ & -887.147 & unstable & -887.147 & 0.075\\
			   \hline\hline
  		\end{tabular}
\end{table}

The 1-ring and 3-ring defect C$_{3:3}$ and C$_{12:12}$ structures considered above present no net permanent magnetic moment, since the spin-polarized edge atoms appear in pair with opposite orientations, resulting in AF ground state with balanced spin-up and spin-down sublattices. However, in a view of spintronic applications, it would be much more interesting to find the GNM structures with nonzero net magnetic moment. This can be actually done by building unbalanced sublattice [Fig.~\ref{fig1}(d)]. We found that the ground state of this unbalanced-defect triangular structure T$_{6:7}$ turns out to be ferrimagnetic (FMi) with total net moment of 3.987~$\mu_B$ per unit cell. This moment originates from each edge atom's dangling bond ($\sigma $-bond) with spin moment of 1~$\mu_B$ providing the magnetic moment of 6(red)-3(blue)=3$\mu_B$ in addition to contribution from $\pi$-bond equal to 1~$\mu_B$ according to total number difference between atoms on A and B sublattice $\Delta _{AB}$. 

\begin{figure}
\begin{center}
\leavevmode
\includegraphics[width=8 cm]{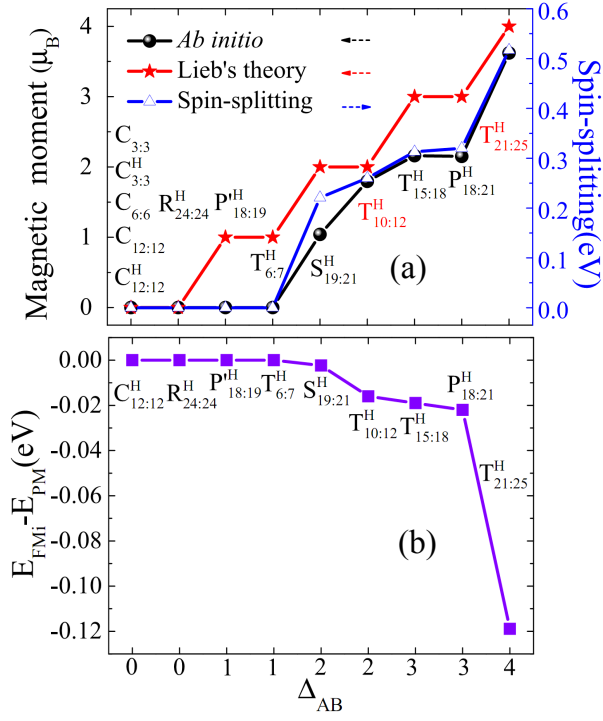}
\caption{(a) Total magnetic moment ($\mu_B$/cell) (left) and spin-splitting (right) as a function of $\Delta_{AB}$ for various GNM geometries, where P$'^H_{18:19}$ is transformed from a pentagon structure (P$^H_{18:21}$) by adding 2 $A$ atoms to the 2 opened hexagons. The result of the Lieb's theorem prediction is also given for comparison. The even number of A+B structures T$^H_{10:12}$ and T$^H_{21:25}$ are shown in red color to indicate the well agreement with Lieb's  theorem prediction. (b) Energy difference between ferrimagnetic and paramagnetic states.}
\label{fig3}
\end{center}
\end{figure}

Dangling bonds at the edge C atoms of non-passivated GNM are strongly chemically reactive\cite{nl}, which leads to hole formation energy of pure GNM higher than 1 eV/C [see~TABLE \ref{table:table I}]. Thus edge C atoms are likely to be passivated by light elements. We used hydrogen for passivation of edge C atoms and considered basic geometrical GNM shapes (Fig.~\ref{fig2}) for which the GNM hole formation energy is dramatically decreased due to passivation of dangling bonds [see~TABLE \ref{table:table I}]. For triangular holes, one can see that the formation energy decays as a function of hole dimension. At the same time, the magnetic moment increases and is roughly proportional to the GNM hole size. When the triangular hole size is increased, one observes that the net moment gets bigger [Fig.~\ref{fig2}(b,c and d)] and reaches 3.62~$\mu_B$ for the biggest hole shown in~Fig.~\ref{fig2}(d). It is interesting to note that Lieb's theorem was originally formulated for even A+B number of atoms, and indeed the obtained values for $T^H_{10:12}$ and T$^H_{21:25}$ follow Lieb's theorem predictions. However, there is an exception for non-triangular case of (S$^H_{19:21}$) with total sum of A and B being even, which is not well accounted by Lieb's theorem. In addition, one can see from~TABLE \ref{table:table I} that the formation energy values of triangular GNMs decrease as a function of hole size and are comparable to those of the non-magnetic configurations.

In Fig.~\ref{fig3}(a) we summarize aforementioned results including the calculated net magnetic moments for circular (C$^H_{12:12}$), rhombic (R$^H_{24:24}$), sector (S$^H_{19:21}$) and pentagon (P$^H_{18:21}$) GNM shapes represented in Fig.~\ref{fig2}(e)-(h), respectively. In addition, the curves contain the net magnetic moment values for alternative pentagon shape GNM, P$'^H_{18:19}$, obtained from P$^H_{18:21}$ by adding 2 A sites to complete 2 hexagons in upper left and upper right 6-rings in Fig.~\ref{fig2}(h). Even though the overall trend of the calculated values qualitatively follows the Lieb's theory, differences are observed, and first-principles calculations do not always correspond to $\Delta_{AB}$, even for the case with A+B is even of S$^H_{19:21}$ as we have already mentioned. In fact, we can ascribe the structures with odd number of A+B atoms, i.e. T$^H_{6:7}$, P$^{'H}_{18:19}$, T$^H_{15:18}$, P$^H_{18:21}$ as well as sector shape S$^H_{19:21}$ GNMs to intermediate regime between nonmagnetic and highly magnetic regimes. This regime provides a root towards design of magnetic GNM supermeshes. It is worth to note that the possible mechanism for the deviation from the Lieb's theorem of the moment value for the sector shape GNM S$^H_{19:21}$ compared to T$^H_{10:12}$ where A+B is even for both, could be attributed to larger amount of armchair edges (not favorable for moment formation) in S$^H_{19:21}$ structure.

\begin{figure}
\begin{center}
\includegraphics[width=8 cm]{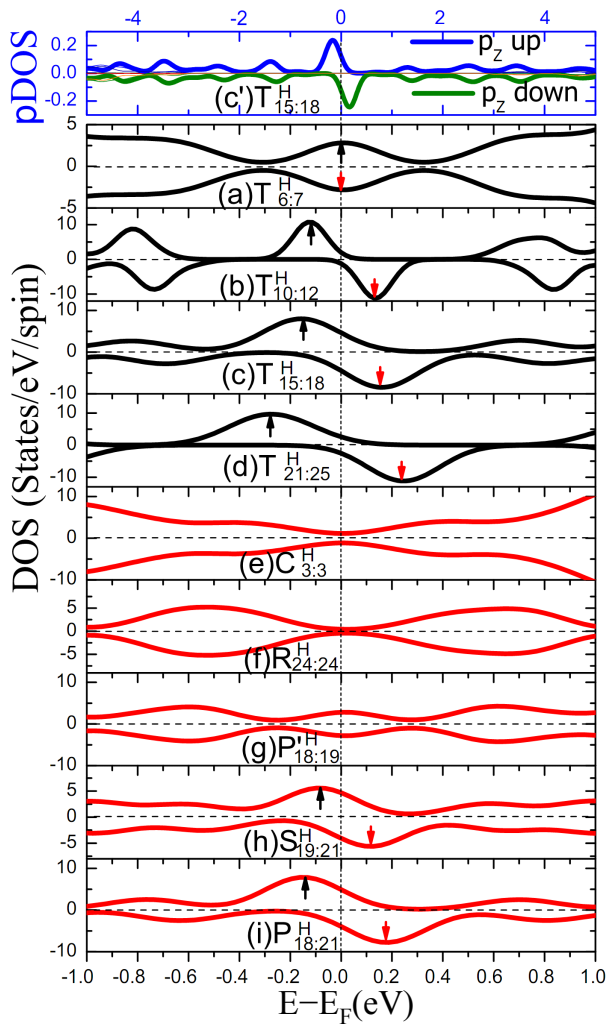}
\caption{Density of states for triangular GNMs of (a) T$^H_{6:7}$, (b) T$^H_{10:12}$, (c) T$^H_{15:18}$, (d) T$^H_{21:25}$, (e) C$^H_{3:3}$, (f) R$^H_{24:24}$, (g)P$^{'H}_{18:19}$, (h) S$^H_{19:21}$, and (i) P$^H_{18:21}$ the peaks around Fermi level are marked with arrows. It can be seen that only p$_z$ state contributes to the moment from (c$'$) projected density of states of one edge atom in the T$^H_{15:18}$ GNM.}
\label{fig4}
\end{center}
\end{figure}

\begin{figure}
\begin{center}
\includegraphics[width=8 cm]{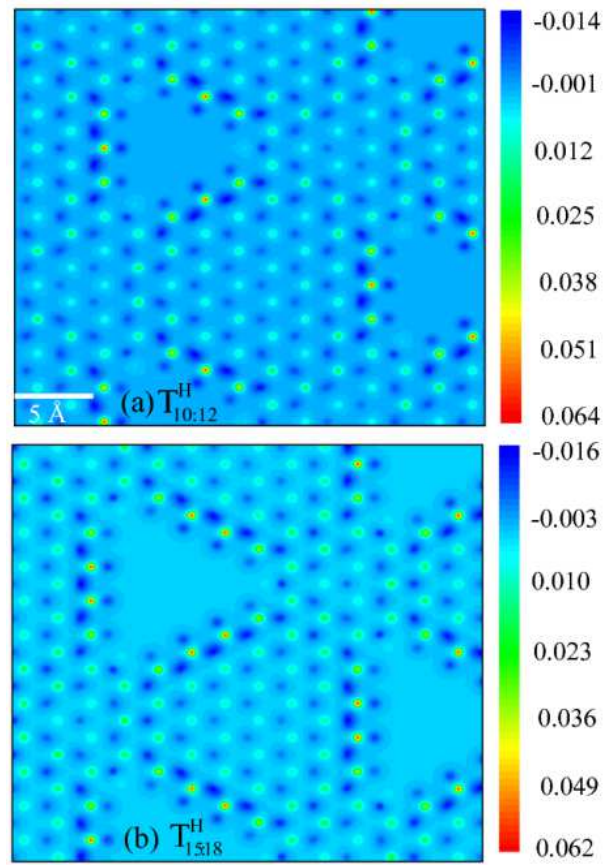}
\caption{Spin density ($\mu_B$/\AA$^3$)~distribution for the two types of graphene nanomesh with localized (upper panel) and delocalized (lower panel) unpaired electrons.}
\label{fig5}
\end{center}
\end{figure}

To further elucidate the origin of magnetism in GNM structures, we compute the total and partial density of states (DOS) for considered GNMs. Fig.~\ref{fig4}(a)-(d) give the total DOS for triangular shape GNMs shown in Fig.~\ref{fig2}(a)-(d), respectively. The exchange splitting between majority and minority spins mainly originates from p$_z$ orbitals, as clearly seen from Fig.~\ref{fig4}(c$^\prime$) where the projected density of states (pDOS) on edge atoms for T$^H_{15:18}$ is plotted. More interestingly, exchange splitting and energy differences between FM and PM states also increase with $\Delta_{AB}$ following the same trend as the net magnetic moment [see Fig.~\ref{fig3}(a) and (b)], reaching values of 0.5~eV and 0.12~eV, respectively. These large exchange splitting values suggest that the magnetism could be preserved at room temperature which look very promising for room temperature graphene spintronics developments. Density of states of structures like T$^H_{10:12}$ provides with the evidence for the presence of localized electrons on the zigzag edges similar to the perfect GNR~\cite{GNRzigzag}. The smearing of the p$_z$ peak and increase of the number of states at the Fermi level (metallization) suggest the electron delocalization on edges and switch from AFM to FM configuration similar to the case of partially oxidized graphite edges~\cite{Kop}. We have plotted the spin density figures for the localized and delocalized cases for illustrate dramatic changes in localization of electrons with vanishing of pseudogap in density of states (see Fig.~\ref{fig5}).The cause of such delocalization of the unpaired electrons on the edges of GNM holes can be attributed to combination of sublattice degeneracy breaking and deviation from the perfect shapes of the graphene nanoribbon.

\section{CONCLUSIONS}

In conclusion, electronic and magnetic properties have been explored in GNM with different geometries using first-principles calculations. For balanced non-passivated GNMs, the ground state was found to be either paramagnetic or antiferromagnetic. In situation of sublattice degeneracy breaking, ferrimagnetic ground states were obtained.
The hydrogen-passivated GNMs were found to be strongly sensitive to the GNM size and shape, with magnetic moments deviation from the Lieb's theorem trend caused by the delocalization of the unpaired electrons on the zigzag edges. The calculations of the formation energy provide with the evidence for the structural stability and high probability of the formation of ferrimagnetic structures. Furthermore, three magnetic regimes are revealed: (i) highly magnetic GNM obeying Lieb's theorem ; (ii) quenched magnetic state due to complete chemical bond reconstruction; and (iii) intermediate regime providing a possible root towards design of magnetic GNM supermeshes. These results demonstrate that a turn from zero dimensional graphene nanoflakes throughout one-dimensional GNR with zigzag edges to GNM breaks localization of unpaired electrons at zigzag edges and provides a deviation from the Lieb's theorem trend. Such delocalization of the electrons allows switching of the ground state of system from antiferromagnetic narrow gap insulator discussed for GNRs to ferromagnetic or nonmagnetic metal. These results combined with obtained large values of the exchange splitting (increasing with $\Delta_{AB}$) pinpoint promising perspectives for developing room-temperature graphene spintronics.

\begin{acknowledgments}
We thank Profs. Xavier Blase, Albert Fert, Irina Grigorieva, Yakov Kopelevich and Oleg Yazyev for fruitful discussions. This work was supported by Chair of Excellence Program of the Nanosciences Foundation in Grenoble, France, by the ANR NANOSIM-GRAPHENE and EU CONCEPT-GRAPHENE.
\end{acknowledgments}

\end{document}